# Kinetics of Charge Transport in Wide-Band Semiconductors at the Detection of X-Ray Radiation

V.Ya. Degoda[*] and A.O. Sofienko

Kyiv National Taras Shevchenko University, Department of Physics
Acad. Glushkova ave., 2, building 1, 03680 Kyiv, Ukraine

As a result of absorption of X-ray quantum in a semiconductor, the generation of electron-hole pairs takes place in a small volume (diameter < 0.5 $\mu$m). Their surplus energy is lost due to the scattering on phonons of the crystal lattice. Spatial distribution of the charge carriers makes the form of current pulse on electrodes of the crystal complicated when an external electric field is applied. We present a logical chart of construction of basic kinetic model of X-ray conductivity (XRC) in semiconductors that uses the successive in time calculation of the spatial distribution of free charge carriers and the diffusive-drift model of motion of free carriers in a solid. The basic form of current pulse in an external circle was obtained in the analytical kind for the case of an ideal semiconductor, e.g. that does not contain deep traps and recombination centers, as well as for the case of a crystal with dominant shallow or deep traps of electrons and holes.

## 1. Introduction

Today, semiconductor materials are most widely used for application in detectors of ionizing radiation. Advantage of semiconductor detectors over other systems of detection (scintillation, track detector) is a direct transformation of energy of ionizing radiation into electric current that allows them to be successfully applicable in spectrometry investigations and applied problems of radiometry [1–3]. However, successful practical application of semiconductor materials in detectors of ionizing radiation is not strengthened by theoretical models of kinetics of radioconductivity. Experimental investigations of X-ray conductivity (XRC) and X-ray luminescence (XRL) of wide-band semiconductors, particularly, of zinc selenide [4–6], indicate significant complexity of the processes caused by detection of ionizing radiation in solids. Moreover, explanation for many experimental facts is impossible in the frame of classical theories of conductivity and charge transport with consideration of the band theory of solids.

We propose to start the construction of kinetic model of radioconductivity of semiconductors with consideration of the as simplified as possible physical picture and to complicate it gradually step by step approaching the real picture. Also, it is reasonable to begin the consideration for ionizing radiation, which does not create new structural defects in the material of detector. Starting from these circumstances, let construct the kinetic model of XRC for an ideal crystal without recombination centers and traps. Then, having the dependence of the form of current pulse on diffusive-drift parameters of motion of carriers, one could do the following step — the consideration of influence of shallow and deep traps on current pulse and collection efficiency of charges and, consequently, on detector response function at the detection of ionizing radiation.

## 2. The X-ray conductivity of an ideal crystal without recombination centers and traps

### 2.1. Model for kinetics of X-ray conductivity

In experimental investigations of XRC and XRL, big number of excitation quanta ($> 10^6$) always takes part. They are inhomogenously absorbed in the material. Absorption of X-ray radiation is determined by the Burger–Lambert law like at photoexcitation that causes macro inhomogeneity of electronic excitations in the crystal. But, unlike photoexcitation, absorption of X-ray quantum is assisted by generation of thousands of charge carriers. That is why the local excitation inhomogeneity occurs at X-ray illumination. For the construction of the basic model of XRC let take into account only basic processes and use the following assumptions:

- absorption of X-ray quantum does not result in creation of new structural defects in material of detector;
- free charge carriers are generated in the very small volume of semiconductor [7] and number of the generated pairs ($N_0$) is determined by energy of X-ray quantum ($h\nu_X$) and by band-gap energy of semiconductor ($E_g$) [8, 9]:

[*] corresponding author; e-mail: `degoda@univ.kiev.ua`

$$N_0 \approx \frac{h\nu_\mathrm{X}}{3E_\mathrm{g}}; \qquad (1)$$

- initial spatial distributions of electrons and holes are identical;
- electric field in a semiconductor is homogeneous and determined by enclosed difference of potentials on electric contacts and by thickness of the sample.

The phenomenological formula (1) obtained experimentally in the investigation of the energy quantum yields X-ray luminescence [9]. It was found that 2/3 of the photoelectron energy is transported into heat (phonons generation) and only 1/3 of the energy goes for the electron-hole pairs generation.

As soon as the rates of different processes of XRC are different, the kinetics of XRC in general case can be divided in three basic stages in time:

1. Generation ($t = 0$–$10^{-12}$ s), when absorption of X-ray quantum and appearance of high-energy electron take place. This high-energy electron generates the initial average spatial distribution of $N_0$ free electrons and $P_0$ free holes ($N_0 = P_0$).

2. Migration ($t = 10^{-11}$–$10^{-6}$ s), when the spatial distribution of free carriers is changing due to their diffusion-drift motion (localization of carriers on the possible recombination centers and traps, delocalization from shallow traps).

3. Relaxation ($t > 10^{-5}$ s), when delocalization of carriers from deep traps occurs and the carriers can be repeatedly localized at the recombination centers or traps, or attain the contacts under the action of the external field with a large time-lag creating in this way constant background current.

The time boundaries of the stages are quite relative and can be changed even by several orders depending on the semiconductor material, the external electric field value and the sample thickness. Such division in the stages allows considerable simplifying of calculations because in every stage, only several processes prevail.

It is possible to apply the following logical chart of construction of the XRC kinetics. At first, let determine the form of current pulse in an external circle at absorption of one X-ray quantum in an ideal semiconductor without recombination centers and traps. This form of pulse will be the start point for the subsequent account of different type of centers, which will change this pulse. Later on, it is necessary to take into account Coulomb interaction between the free carriers of the opposite charge. After this, it is possible to enter the point defects in material — firstly, shallow traps, then to carry on with consideration of deep traps and recombination centers.

Introduction of point defects in the calculation system will allow to obtain kinetics of change of the spatial distributions of free carriers and so, to obtain change of the form of current pulse. Having a form of current pulse at absorption of one X-ray quantum, it is possible to estimate the charge collection efficiency. Summation of individual pulses will allow to determine the general current of XRC.

### 2.2. Generation and thermalization of free charge carriers

At interaction of X-ray quantum of energy $1 \div 50$ keV with a solid, the basic processes are absorption, ionization of ion and appearance of high-energy electron. This electron creates on the average $N_0$ of electron–hole pairs in a local volume due to the ionization losses of kinetic energy in thermalization process ($10^{-13}$–$10^{-12}$ s). Applying the diffusion model of thermalization of electron, it is possible to get the average spatial distribution of concentration of generated free charge carriers $N_0(r)$, which is well described by the Gaussian distribution [7]:

$$N_0(r) = \frac{N_0}{(2\pi)^{\frac{3}{2}} r_\mathrm{g}^3} \exp\left(-\frac{r^2}{2r_\mathrm{g}^2}\right),$$

$$\text{where} \quad r_\mathrm{g} = \frac{6\pi\varepsilon_0^2 \sqrt{E_\mathrm{g}(h\nu_\mathrm{X})^3}}{e^4 n_\mathrm{e} \ln(h\nu_\mathrm{X}/3E_\mathrm{g})}, \qquad (2)$$

where $n_\mathrm{e}$ is the concentration of electrons in semiconductor materials, $r_\mathrm{g}$ is the average radius of the spatial distribution of carriers, $r$ is a distance from the center of the distribution of carriers. Such spatial distribution of free electrons and holes is the initial for the next migration stage.

Generated charge carriers become thermal for the time $\approx 10^{-12}$–$10^{-11}$ s due to interaction with lattice phonons. An analysis shows [10] that a dominant process is the scattering on optical nonpolar phonons. So, for example, at initial energy of carriers 1–10 eV in Ge, the specific energy losses at scattering on optical phonons are 20 times higher than at scattering on acoustic phonons. At the scattering on optical nonpolar phonons, when $W^\pm(t) \gg k\theta$ ($\theta$ is the Debye temperature), the dynamics of energy losses is described by relation [10–11]:

$$\left(-\frac{\mathrm{d}W^\pm}{\mathrm{d}t}\right)_\mathrm{DO}$$
$$= \frac{(m^\pm)^{3/2} \varXi^2 \sqrt{3}}{2\pi^{3/2}\hbar^2 \rho} \frac{k\theta}{\sqrt{W^\pm}} \mathrm{e}^{-\frac{3k\theta}{4W^\pm}} K_1\left(\frac{3k\theta}{4W^\pm}\right), \quad (3)$$

where $m^\pm$ is the effective mass of carriers, $\varXi$ is a constant of optical deformation potential [10, 11], $\rho$ is a density of crystal matter, $K_1(x)$ is the first-order Bessel function of the second kind, which can be expressed as [10]:

$$K_1(x) = x \int_1^\infty \mathrm{e}^{-xt} \sqrt{t^2-1}\,\mathrm{d}t \approx \frac{1}{x}\mathrm{e}^{-\frac{x}{2}};$$

$$10^{-2} \le x < 1.$$

Carriers of initial energy $W \gg kT$ in solid can be considered as being free; therefore dependence of diffusion coefficients of nonequilibrium "hot" charge carriers can be determined by relation

$$D^{\pm}(t) = \frac{1}{3}v^{\pm}(t)l^{\pm}(t), \tag{4}$$

where $v^{\pm}(t) = \sqrt{2W^{\pm}(t)/m^{\pm}}$ is a velocity of charge carriers, $l^{\pm} = v^{\pm}(t)/[(\tau_{\text{AC}}^{-1})^{\pm} + (\tau_{\text{DO}}^{-1})^{\pm}]$ is a mean free path determined by the scattering processes, $\tau_{\text{DO}}$ is the relaxation time at the scattering on optical nonpolar phonons during thermalization, $\tau_{\text{AC}}$ is the relaxation time at the scattering of already thermal equilibrium carriers on acoustic phonons. In case, when the scattering on optical phonons prevails at thermalization and after thermalization, the scattering of carriers can be considered taking into account only acoustic phonons, the relation (4) can be expressed as a function of carrier energy and total relaxation time

$$D^{\pm}(t) = \frac{2}{3}\frac{W^{\pm}(t)}{m^{\pm}\left[(\tau_{\text{AC}}^{-1})^{\pm} + \tau_{\text{DO}}^{-1}(W^{\pm}(t))\right]} \to D_0. \tag{5}$$

Estimation of relation (5) taking into account (3) allows to obtain dependences of the energy and diffusion coefficients of carriers on their thermalization time

$$W^{\pm}(t) \approx \frac{3}{2}kT + \left(e^{\frac{k\theta}{W_0}}\sqrt{W_0} - tC_{\text{DO}}^{\pm}\right)^2, \tag{6}$$

$$D^{\pm}(t) = \frac{2}{3}\frac{W(t)}{m^{\pm}\left[(\tau_{\text{AC}}^{-1})^{\pm} + 2\frac{C_{\text{DO}}^{\pm}}{k\theta}e^{-\frac{k\theta}{W(t)}}\sqrt{W(t)}\right]},$$

$$C_{\text{DO}}^{\pm} = \frac{(m^{\pm})^{3/2}\Xi^2}{\pi^{3/2}\hbar^2\rho\sqrt{3}}, \tag{7}$$

where $W_0$ is the initial kinetic energy of photoelectron. Thermalization time ($T_{\text{term}}$), during which the carrier energy becomes equal to the equilibrium thermal energy of the crystal lattice, is $T_{\text{term}} = C_{\text{DO}}^{-1}\exp(k\theta/W_0)\sqrt{W_0}$. Calculating (6), one can choose the initial value of the energy of "hot" carriers of the same order as the band gap energy of the semiconductor ($W_0 \approx E_g$). Indeed, this value can be considered as boundary one, when secondary ionization of the media turns out to be impossible (for generation of one electron–hole pair, the energy of $(2.5-3) \times E_g$ upon the average is used [9, 12]). Considering that motion of electrons and holes during the thermalization process is diffusive motion with variable diffusion coefficients $D^{\pm}(t)$ and the influence of external electric field is inconsiderable, because $W^{\pm} \gg |eE_0l|$, the spatial distribution of carriers at any time moment will be described by Gaussian function

$$N^{\pm}(r,t) = \frac{N_0}{\left[2\pi\left(r_g^2 + \sigma^2(t)\right)\right]^{\frac{3}{2}}}e^{-\frac{r^2}{2r_g^2+2\sigma^2(t)}}, \tag{8}$$

where $\sigma(t)$ is the average radius, the spatial distribution of carriers at thermalization at time moment $t$ enlarged in. The increase of dispersion of carriers distribution during time d$t$ at the time moment $t$ owing to diffusion is $d(\sigma^2(t)) = D(t)dt$. Calculating the dependence (5), we obtain an average sphere radius $R_{\text{term}}$ of the spatial distribution of charge carriers after thermalization

$$R_{\text{term}} = \sqrt{r_g^2 + \int_0^{T_{\text{term}}} D(t)\,dt}.$$

Hereby, the spatial distribution of electrons and holes after thermalization is described by Gaussian function, as well as after generation stage [7], but in Eq. (2), it is necessary to make a substitution $r_g^2 \to (R_{\text{term}})^2$. In fact, such substitution means that it is possible to combine the generation and thermalization stages in one getting corresponding parameters of final spatial distribution of generated carriers, which will be initial distribution for the migration stage. Relation (8) allows calculating of self-field of charge carriers as well as its influence on the kinetics of carrier drift, and so on the formation of rising edge of pulse of conduction current created in the external electric circle of detection system. Note that numerical estimations following the relations (6)–(8) point to considerable reduction (to $\approx$ 100 V/cm) of self-field of carriers during 100–1000 ps in different materials.

### 2.3. The spatial distribution of carriers at drift

If there is an external electric field and a gradient of concentration of generated charge carriers in the crystal, which does not contain recombination and localization centers (for taking into account these centers it is necessary to use general system of the kinetic charge transport equations [13]), there are drift and diffusion currents, whose densities in general case are defined by relation

$$\boldsymbol{J}^{\pm} = eN^{\pm}(x,y,z,t)\mu^{\pm}\boldsymbol{E} \mp eD^{\pm}\boldsymbol{\nabla}N^{\pm}(x,y,z,t). \tag{9}$$

Sign "+" in (3) is attributed to the holes, sign "−" — to the electrons. $N^+(x,y,z,t)$ and $N^-(x,y,z,t)$ are the spatial distributions of concentration of generated free holes and electrons, $\boldsymbol{E}$ is a vector of electric field, $\mu^{\pm}$ is a mobility of carriers. In order to have the certain direction of the carriers drift we use the Cartesian coordinate system, where the OX axis opposite to the direction of the vector of external electric field. In accepted coordinate system, relation (9) supplemented with continuity equation, which defines time changes of concentration of carriers, allows to obtain the system of kinetic equations of diffusion-drift motion of electrons and holes

$$\frac{dN^-(x,y,z,t)}{dt}$$
$$= D^-\Delta N^-(x,y,z,t) - \mu^-\boldsymbol{E}\cdot\boldsymbol{\nabla}N^-(x,y,z,t),$$

$$\frac{dN^+(x,y,z,t)}{dt}$$
$$= D^+\Delta N^+(x,y,z,t) + \mu^+\boldsymbol{E}\cdot\boldsymbol{\nabla}N^+(x,y,z,t). \tag{10}$$

It is supposed that size of the local area of generation of free charge carriers is much smaller than the thickness of detector $(d)$ and transversal sizes of detector (along directions $OY$ and $OZ$) highly exceed the thickness. This allows to use delta-function approximation as initial distribution of carriers at the solution of system (10) (because $r_g$ in (2) is much smaller than size of the drift area in semiconductor detector). When the carriers reach the electrodes, they disappear from total distribution, so for solution of equation system (10), the following boundary conditions must be used:

$$N^+(0, y, z, t) = N^-(0, y, z, t) = 0,$$

$$N^+(d, y, z, t) = N^-(d, y, z, t) = 0. \quad (11)$$

Complete solution of equations (10) defines the spatial distribution of concentration of carriers at drift and could be obtained by means of variable separation method

$$N^\pm(x, y, z, t) = \frac{N_0}{4\pi D^\pm t} \exp\left(-\frac{(y-y_0)^2 + (z-z_0)^2}{4D^\pm t}\right.$$
$$\left. \pm \frac{\mu^\pm E(x - x_0)}{2D^\pm} - \frac{(\mu^\pm E)^2 t}{4D^\pm}\right) \frac{2}{d} \sum_{n=1}^{\infty}$$
$$\left(\exp\left(-\left(\frac{\pi n}{d}\right)^2 D^\pm t\right) \sin\left(\frac{\pi n x}{d}\right) \sin\left(\frac{\pi n x_0}{d}\right)\right), \quad (12)$$

where $x_0$ is a coordinate of absorption X-ray quantum in the semiconductor. The calculated distribution functions of concentration of free electrons and holes along direction $OX$ at the different time moments in Si semiconductor detector are given in Fig. 1. For the calculation, the data for mobility of charge carriers in Si from [14] were used.

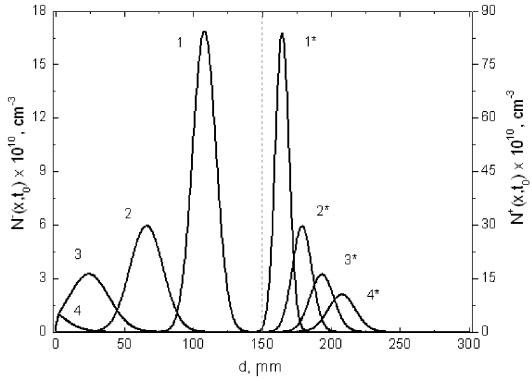

Fig. 1. Distribution functions of concentration of free electrons and holes (*) along direction of the OX axis (from left to right) at the different time moments in Si semiconductor detector: 1, 1* — 10 ns; 2, 2* — 20 ns; 3, 3* — 30 ns; 4, 4* — 40 ns ($d = 300$ μm, $x_0 = 150$ μm, $E = 300$ V/cm, $T = 300$ K, $h\nu_X = 5$ keV).

Figure 1 shows that the concentrations of electrons and holes in the drift differ by several times. This is due to the fact that the carrier mobility and, consequently, their diffusion coefficients are also different in several times.

### 2.4. Form of current pulse of XRC

For the calculation of current in an external electric circle $i(t)$, that free charge carriers create at drift to the electrodes in the electric field, it is necessary to apply the Ramo–Shockley theorem for point charges [15, 16]. Whereas the work of electric field on displacement of all free carriers is determined as a sum of works on moving of each carrier, the value of current in an external circle is determined as a sum of currents created by all free charge carriers

$$i(t) = \frac{q^-(t)\mu^- E}{d} + \frac{q^+(t)\mu^+ E}{d}$$
$$= \frac{eN_0 E}{d}\left[E(t)\mu^- + P(t)\mu^+\right]. \quad (13)$$

Introduction of $E(t)$ and $P(t)$ functions (the part of generated electrons and holes, which are staying free at the time moment $t$ and continuing the drift) is necessary for subsequent taking into account the processes of carrier localization on traps and recombination centers.

$$E(t) = \frac{1}{N_0} \int_{-\infty}^{\infty} \int_{-\infty}^{\infty} \int_0^d N^-(x, y, z, t)\,dx\,dy\,dz,$$

$$P(t) = \frac{1}{N_0} \int_{-\infty}^{\infty} \int_{-\infty}^{\infty} \int_0^d N^+(x, y, z, t)\,dx\,dy\,dz. \quad (14)$$

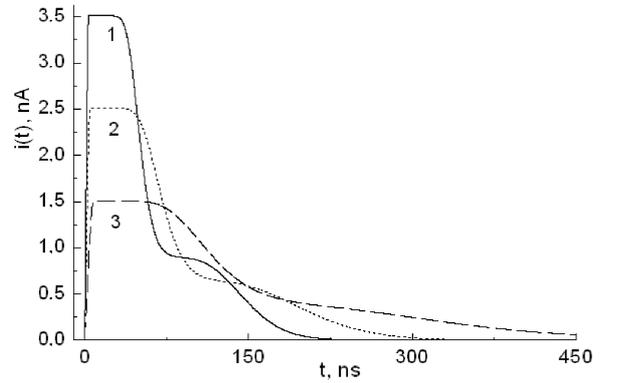

Fig. 2. Current pulses in Si detector at different value of electric field: 1 — 140 V/cm, 2 — 100 V/cm, 3 — 60 V/cm ($d = 200$ μm, $x_0 = 100$ μm, $T = 300$ K, $E_X = 5$ keV).

The calculation of (14) is considerably complicated by the presence of infinite series in the exact solution relatively to carrier concentration function (12). The detailed analysis makes it possible to derive the simple analytical dependences for their approximation

$$E(t) = \frac{1}{1 + \exp\left(\frac{\mu^- E t - x_0}{\sqrt{D^- t/2}}\right)};$$

$$P(t) = \frac{1}{1 + \exp\left(\frac{\mu^+ E t - (d - x_0)}{\sqrt{D^+ t/2}}\right)}. \quad (15)$$

Formulae (15) are obtained in the approximation of the exact relations (14) for different values of the mobility of charge carriers and the electric field in different materials (ZnSe, Ge, Si, CdTe).

The use of these approximating relations allows to simplify considerably the calculations and gives analytical functions of current pulse keeping dependence on the basic kinetic parameters of carrier motion. Figure 2 shows the calculations of the current pulse in Si for different values of the external electric field.

## 3. XRC in the presence of localization centers in the crystal

### 3.1. Ratio between free and localized carriers

At initial time moment, all generated carriers of charge are in free state. In the process of their diffusive drift motion in the crystal lattice, they begin to be localized on shallow traps. If lifetime of carriers relatively to localization processes ($\tau_0$) is much shorter than drift time, dynamic equilibrium between number of free ($N^\pm$) and localized ($N_0 - N^\pm$) carriers sets in after some time. This process is described by the kinetic equation

$$\frac{dN^\pm}{dt} = -\frac{N^\pm}{\tau_0^\pm} + \frac{N_0 - N^\pm}{\tau_i^\pm} \to N^\pm(t)$$

$$= \frac{N_0}{1+\frac{\tau_i^\pm}{\tau_0^\pm}}\left(1 + \frac{\tau_i^\pm}{\tau_0^\pm}\exp\left(-\frac{t}{\tau_0^\pm}\right)\right). \quad (16)$$

Relation (16) determines number of free charge carriers, which are in drift and create electrical current in external circle of detector. $N^\pm(t)/N_0$ ratio is a probability to be in free state for every electron, regardless of his position in space.

### 3.2. Probability of frequent localization of charge carriers

If the drift time and average lifetime in free state are known, it is possible to find the average number of successive localizations of carriers $m_0$ at drift $m_0^\pm = (T_{\rm dr}^\pm/\tau_0^\pm) - 1$. The average number of successive localizations is not necessarily an integer number, because the localization process is a statistical process. Considering multiple successive localizations, an important task is the determination of probability that carriers will be trapped at fixed number of time during the drift time.

The account that the acts of localization are processes independent of each other results in probability of localizations as Poisson distribution $F_{\rm m}(m_0) = m_0^m e^{-m_0}/m!$ (when the average number of localizations $m_0 \geq 20$, one can use the Gauss distribution). This allows to determine the number of carriers localized exactly $m$ during the drift: $N_{0m} = N_0 F_{\rm m}(m_0)$. The whole process of collecting of $N_0$ carriers on the electrode could be divided in $m_{\max}$ independent processes

$$m_{\max} = m_0\left[1 + \sqrt{\frac{1}{m_0}\ln\left(\frac{N_0^2}{2\pi m_0}\right)}\right].$$

For each $m$ group of carriers, the time in a detector will be increased to the mean value of the total time of localization on traps $\Delta t_{\rm m} = m\tau_i$. So, duration of the current pulse in the external circle should be considered as $T_{\rm dr} + m_{\max}\tau_i$.

### 3.3. Kinetics of drift of free charge carriers at localization on traps

At localization on traps m times, the average time of carriers in free state is not $\tau_0^\pm$, but $\tau_{0m}^\pm = T_{\rm dr}^\pm/(m+1)$. Correspondingly, the kinetic equation (7) for m group of carriers becomes

$$\frac{dN_m^\pm(t)}{dt} = -\frac{(m+1)N_m^\pm(t)}{T_{\rm dr}^\pm} + \frac{N_{0m} - N_m^\pm(t)}{\tau_i^\pm}. \quad (17)$$

Correspondingly, solution of Eq. (17) is

$$N_m^\pm(t) = N_0 F_{\rm m}(m_0)/(1 + (m+1)\tau_i^\pm/T_{\rm dr}^\pm)$$

$$\times \left(1 + \frac{(m+1)\tau_i^\pm}{T_{\rm dr}^\pm}\exp\left(-\frac{(m+1)t}{T_{\rm dr}^\pm}\right)\right). \quad (18)$$

$$F_{\rm m}(m_0) = e^{-(m-m_0)^2/(2m_0)}/\sqrt{2\pi m_0}$$

In order to take into account progressive collecting of charge on the electrodes of detector as well as reduction of charge carriers in the crystal volume, it is necessary to consider diffusive motion of carriers at drift. In an ideal semiconductor, the carriers reach the corresponding electrode not simultaneously as a result of diffusion. The distribution function of number of drifting electrons over the time is determined through integration of their spatial distribution function within coordinates of detector (15). The product $D \cdot t$ in denominator of exponent in (15) must be substituted by $D \cdot T_{\rm dr}$ at consideration of localization of carriers, because localization on traps does not change diffusive expansion of spatial distribution of carriers, but creates their delay in the crystal at drift and frames considerable expansion of current pulse over the time.

At the presence of intermediate localizations of carriers on traps, uncertainty of getting corresponding electrode is additionally increased as a result of statistical behavior of time of being in localized state. Diffusion motion results in dispersion of spatial distribution of carriers $2DT_{\rm dr}$. An equivalent time period of collecting of carriers on electrode is $\sqrt{2DT_{\rm dr}}/\mu E_0$. Process of delocalization of carriers from traps is a statistical process and probability of being in localized state is determined as $w_i(t) = \exp(-t/\tau_i)/\tau_i$. Hence, dispersion of localization on one shallow trap over the time is

$$D(t) = \frac{1}{\tau_i}\int_0^\infty (t - \tau_i)^2\exp\left(-\frac{t}{\tau_i}\right)dt = \tau_i^2.$$

As far as the process of successive $m$ localizations at drift is considered, the total dispersion for $m$ independent of each other localization processes will be $D(\sum_j^m t_j) = \sum_j^m D(t_j) = m\tau_i^2$. Delay in time of getting $N_{\rm m}$ carriers to electrode must be taken into account as well: $(T_{\rm dr} + m\tau_i)$. So, we obtain that at localization of carriers on shallow traps, relation (15), for example, for electrons will be changed as following

$$E_{\rm m}(t, x_0, m)$$

$$= \left(1 + \exp\left(\frac{t - m\tau_i^- - T_{\rm dr}^-}{\sqrt{\frac{D \cdot T_{\rm dr}^-}{2(\mu^- E_0)^2} + m\left(\tau_i^-\right)^2}}\right)\right)^{-1}. \quad (19)$$

The same result is supposed for holes at localization on shallow traps. Using (18, 19) we obtain that total probability of carriers being in drift is determined by the sum of products of probabilities of separated and independent processes of localization and diffusion-drift motion

$$N^-(t) = N_0 \sum_m \left( \frac{F_m(m_0^-)E_m(t,x_0,m)}{1+(m+1)\tau_i^-/T_{dr}^-} \right.$$
$$\left. \times \left[1 + \frac{(m+1)\tau_i^-}{T_{dr}^-} \exp\left(-\frac{(m+1)t}{T_{dr}^-}\right)\right]\right),$$
$$N^+(t) = N_0 \sum_m \left( \frac{F_m(m_0^+)P_m(t,x_0,m)}{1+(m+1)\tau_i^+/T_{dr}^+} \right.$$
$$\left. \times \left[1 + +\frac{(m+1)\tau_i^+}{T_{dr}^+} \exp\left(-\frac{(m+1)t}{T_{dr}^+}\right)\right]\right). \quad (20)$$

As a limit approximation of low concentration of traps, when $m_0^\pm \to 0$, the current pulse calculated correspondingly to Ramo–Shokli theorem with using relation (20) will asymptotically tend to the current pulse functions (13) for motion of carriers in an ideal crystal. Calculation of electron component of current pulse in Si and ZnSe at the presence of shallow traps is given in Fig. 3.

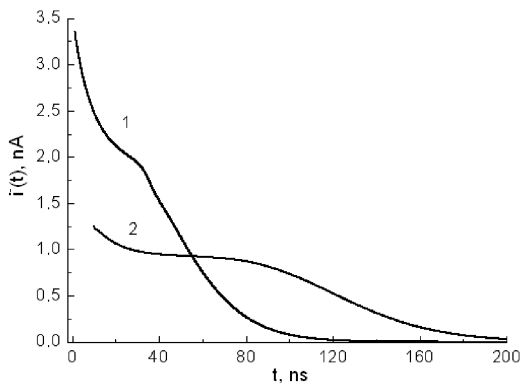

Fig. 3. Calculated electron component of current pulse in the crystal: 1 — Si, 2 — ZnSe ($d = 300$ $\mu$m, $x_0 = 150$ $\mu$m, $E_0 = 300$ V/cm, $E_X = 5$ keV, $\tau_i^- = \tau_0^- = 10$ ns).

Results of calculations show that different values of carrier mobility in referred materials cause remarkable difference in the number of intermediate localizations during drift at the same parameters of traps, and thus the form of XRC current pulse changes.

## 4. Conclusions

Presented model of kinetics of XRC of semiconductors allows to get the form of current pulse at absorption of X-ray quantum in an ideal semiconductor, and to analyze influence of basic parameters of material and value of electric field on it. Analytical relations for current pulse, that significantly simplify the subsequent development of kinetic model of XRC for semiconductors, are proposed. The influence of shallow traps on the current pulse of XRC is considered. The influence of the average number of localization of carriers at drift on the pulse form in different materials is analyzed as well.

The obtained relations make it possible to get the quantitative data for the charge response of semiconductor detectors depending on characteristics of a given material, radiation parameters and the external electric field value. The theoretical analysis of the drift and diffusion of carriers is necessary to determine the parameters of the current pulse in a thick semiconductor detector ($\approx$ 1–5 mm).

### Acknowledgments

The work was executed at sponsorship of Fundamental Researches State Fund of Ukraine (project 25.4/138).